\documentclass[aps,prd,superscriptaddress,twocolumn,floatfix,nofootinbib]{revtex4-1}
\usepackage{epsfig,bm}
\usepackage{graphicx,amssymb}
\usepackage{amsmath}
\usepackage{physics}
\usepackage{tensor}
\usepackage{diagbox}
\usepackage{epstopdf}
\usepackage{float}
\usepackage{tabularx}
\usepackage{subfigure}
\usepackage{soul}
\usepackage{tikz-feynman}
\tikzfeynmanset{compat=1.0.0}

\usepackage{soul}

\usepackage[eps]{pstricks}
\usepackage[normalem]{ulem}  %
\renewcommand{\sout}{\bgroup \color{red} \ULdepth=-.5ex \ULset}

\begin{document}


\title{Production and in-medium modification of $\phi$ mesons in proton-nucleus reactions from a transport approach}



\author{Philipp Gubler}
\email[]{philipp.gubler1@gmail.com}
\affiliation{Advanced Science Research Center, Japan Atomic
Energy Agency, Tokai, Ibaraki 319-1195, Japan}

\author{Masaya Ichikawa}
\email[]{michikaw@rcnp.osaka-u.ac.jp}
\affiliation{Advanced Science Research Center, Japan Atomic
Energy Agency, Tokai, Ibaraki 319-1195, Japan}
\affiliation{Institute of Particle and Nuclear Studies, High Energy Accelerator Research Organization (KEK), Tsukuba 305-0801, Japan}
\affiliation{RIKEN Nishina Center, 2-1 Hirosawa, Wako, Saitama 351-0198, Japan}

\author{Taesoo Song}
\email[]{T.Song@gsi.de}
\affiliation{GSI Helmholtzzentrum f{\"u}r Schwerionenforschung GmbH,
Planckstrasse 1, D-64291 Darmstadt, Germany}

\author{Elena Bratkovskaya}
\email[]{E.Bratkovskaya@gsi.de}
\affiliation{GSI Helmholtzzentrum f{\"u}r Schwerionenforschung GmbH,
Planckstrasse 1, D-64291 Darmstadt, Germany}
\affiliation{Institut f{\"u}r Theoretische Physik, Johann Wolfgang Goethe-Universit{\"a}t,
Max-von-Laue-Str. 1, D-60438 Frankfurt am Main, Germany}
\affiliation{Helmholtz Research Academy Hesse for FAIR (HFHF),
GSI Helmholtz Center for Heavy Ion Physics, Campus Frankfurt, 60438 Frankfurt, Germany}


\date{\today}

\begin{abstract}
Production and in-medium modification of hidden strange $\phi$ mesons are studied in proton-nucleus reactions - p$+$C, p$+$Cu and p$+$Pb - at 12 GeV$/c$, partially measured by the KEK E325 collaboration via the dilepton decay mode. 
This work is based on the off-shell microscopic Parton-Hadron-String Dynamics (PHSD) transport approach, 
 which provides the dynamical description of strongly interacting hadronic and partonic degrees of freedom in dense matter, and especially makes it possible to simulate in-medium off-shell dynamics of the $\phi$ meson in the reactions studied. 
Different in-medium scenarios for the modification of the $\phi$ meson spectral function in nuclear matter are investigated:  
i) dropping pole mass, ii) collisional broadening and iii) simultaneous dropping pole mass and collisional broadening. 
Even though the $\phi$ meson  production in p+A reactions occurs only at densities around or below normal nuclear matter density $\rho_0$, we find visible modifications of the generated dilepton spectra for the in-medium scenarios compared to the vacuum distribution. 
\end{abstract}


\maketitle


\section{Introduction \label{sec:Intro}}

Even though the interaction of the $\phi$ meson with a nucleon and its ensuing in-medium modification in nuclear matter have
been studied for decades, they are still far from being completely understood. Early attempts estimated 
the finite density behavior of the $\phi$ in QCD sum rule approaches \cite{Hatsuda:1991ez} and 
chiral effective theories \cite{Ko:1992tp} and, independently, the $\phi$-N interaction 
through the QCD van der Waals force, mediated by gluon exchanges \cite{Gao:2000az}. 
All these approaches obtained attractive $\phi$-N or $\phi$-nucleus interactions, which 
pointed to the possibility of a $\phi$-N bound state \cite{Gao:2000az} or were alternatively 
interpreted as an in-medium mass shift of the $\phi$ meson in nuclear matter \cite{Hatsuda:1991ez}. 
While QCD sum rules and the QCD van der Waals force based approach focused only on the real part 
of the interaction, chiral effective theory emphasized the importance of absorption 
processes of the $\phi$ in nuclear matter, leading to a strong increase 
of its decay width. Some of these theoretical predictions have been further refined 
and updated over the years (see e.g. Refs.\,\cite{Asakawa:1994tp,Klingl:1997kf,Klingl:1997tm,Lee:1997zta,Oset:2000eg,Cabrera:2002hc,Zschocke:2002mp,Gubler:2015yna,Gubler:2016itj,Cabrera:2016rnc,Cabrera:2017agk,Cobos-Martinez:2017vtr,Cobos-Martinez:2017woo,Kim:2019ybi,Kim:2022eku}), but the 
basic qualitative conclusions remained unchanged. 

The first experimental results about the behavior of the $\phi$ meson in 
nuclear matter were reported from SPring-8, KEK, CLAS and COSY-ANKE in the first decade of the 
millennium \cite{Ishikawa:2004id,KEK-PS-E325:2005wbm,E325:2006ioe,CLAS:2010pxs,Polyanskiy:2010tj,Hartmann:2012ia}. 
Among them, the result of the KEK E325 experiment is especially noteworthy, as 
it was the only one that constrained both mass shift and in-medium decay width of the $\phi$ meson in 
nuclear matter, reporting a negative mass shift of $-35 \pm 7$ MeV and a broadening factor of 
$3.6^{+1.8}_{-1.2}$ at normal nuclear matter density \cite{KEK-PS-E325:2005wbm}. 

This topic recently received renewed attention thanks to new experimental results from HADES \cite{HADES:2018qkj}, CLAS \cite{Strakovsky:2020uqs} and ALICE \cite{ALICE:2021cpv}, the theoretical interpretation of the HADES data \cite{Paryev:2022zkt} and results from lattice QCD \cite{Lyu:2022imf}, which in total may alter our previous understanding of the $\phi$-N \cite{Chizzali:2022pjd,Sun:2022cxf,Kuros:2024dhc} and $\phi$-nucleus interactions \cite{Filikhin:2024avj,Filikhin:2024xkb}. Specifically, through a femtoscopy measurement, ALICE was able to constrain both real and imaginary parts of the $\phi$-N scattering length $a_{\phi N}$, and reported a large (and attractive) real part, $\mathrm{Re}(a_{\phi N}) = 0.85 \pm 0.34\,(\mathrm{stat.})\pm 0.14\, (\mathrm{syst.})$ fm, and a small imaginary part almost consistent with zero, $\mathrm{Im}(a_{\phi N}) = 0.16 \pm 0.10\,(\mathrm{stat.})\pm 0.09\, (\mathrm{syst.})$ fm (see, however, the more recent analysis in Ref.\,\cite{Feijoo:2024bvn} of the same data, that obtained a qualitatively different conclusion). Note that these values constitute an average of the spin 1/2 ($a_{\phi N}^{1/2}$) and spin 3/2 ($a_{\phi N}^{3/2}$) values. 
On the other hand, the lattice HAL QCD Collaboration computed $a_{\phi N}^{3/2}$ directly from QCD, obtaining a value of $\mathrm{Re}(a_{\phi N}^{3/2}) = 1.43 \pm 0.23\,(\mathrm{stat.})\pm 0.14\, (\mathrm{syst.})$ fm \cite{Lyu:2022imf}. 
It is interesting to observe that these rather large values of the scattering length are in stark disagreement with its absolute value obtained from the $\phi$ meson photoproduction data at CLAS, which gives
$|a_{\phi N}| = 0.063 \pm 0.010$ fm \cite{Strakovsky:2020uqs} and is therefore at least an order of magnitude smaller. 
This discrepancy is not yet well understood. 

It is - in this context - instructive to relate the $\phi$-N scattering length to the potential depth (or mass shift) of the $\phi$ meson in nuclear matter, which can be done qualitatively in the linear density approximation, giving (see for example Ref.\,\cite{Paryev:2022zkt} and the references cited therein for more detailed discussions) 
\begin{align}
V(\rho) = - \frac{2\pi}{m_{\phi}} \Bigl(1 + \frac{m_{\phi}}{m_N}\Bigr) a_{\phi N} \rho, 
\label{eq:sc_len_opt_pot}
\end{align}
which can be rewritten as
\begin{align}
V(\rho) = - 85 \, \Bigl(\frac{a_{\phi N}}{\mathrm{fm}}\Bigr) \frac{\rho}{\rho_0}\,\mathrm{MeV}, 
\label{eq:sc_len_opt_pot_2}
\end{align}
showing that the ALICE measurement corresponds to a potential depth of about 70 MeV, much larger 
than the mass shift reported by the KEK E325 experiment. As discussed in Ref.\,\cite{Paryev:2022zkt}, 
Pauli corrections to the linear density expression are significant and reduce the potential depth by 
a factor of 2, eventually leading to a rather good agreement with the KEK E325 result. 
It will be interesting to see whether this qualitative agreement persists once more precise data 
both from femtoscopy measurements and dileptons from pA collisions become available \cite{Aoki:2023qgl}.   

From an experimental point of view, it is generally a challenging task to measure $\phi$ mesons, 
because of its suppressed production compared to other light vector mesons ($\rho$ and $\omega$). 
While the dilepton decay $\phi \to e^+e^-$ provides a clear signal, as dileptons are not 
affected by the strong interaction, the strong decay mode $\phi \to K^+K^-$ may also be 
explored \cite{KEK:Sako2022} and could be especially useful for measuring the polarization 
of the $\phi$ in proton-nucleus (pA) reactions \cite{Park:2022ayr}. 

Numerous experimental efforts have been undertaken to ascertain the in-medium characteristics of vector mesons in pA and nucleus-nucleus (AA) collisions, primarily through the analysis of dilepton decay channels. Concurrently, theoretical models have been developed to interpret these findings (see Ref.\,\cite{Linnyk:2015rco,Bleicher:2022kcu} for recent reviews). 
However, the focus of these studies has predominantly been on the $\rho$ and $\omega$ mesons, such as for instance 
Ref. \cite{Bratkovskaya:2001ce}, where the dropping mass and collisional broadening scenarios of the $\rho$ and $\omega$ meson spectral functions in pA reactions were studied within the BUU approach in comparison to the dilepton measurement by the KEK-PS E325 collaboration \cite{E325:2000smw}. However, the lack of precision of the experimental data for such broad resonances did not allow 
for any robust conclusions. 
The $\phi$ meson presents a greater challenge due to its infrequent production and relatively narrow width, complicating the study of its in-medium properties. 

In a recent study \cite{Song:2022jcj}, two of the present authors investigated $\phi$-meson production and its relation to in-medium strangeness production (involving kaons and antikaons) in heavy-ion collisions. The analysis spanned a wide energy range, from subthreshold energies ($\sim$ 1 A GeV) to relativistic energies ($\sim$ 21 A TeV), using the off-shell Parton-Hadron-String Dynamics (PHSD) approach. 
Novel meson-baryon and meson-hyperon production channels specifically for $\phi$ mesons were  included, which were calculated using a T-matrix coupled channel approach based on an extended SU(6) chiral effective Lagrangian. The collisional broadening effect on the $\phi$ meson spectral function within the medium was also considered. 
The obtained results exhibited a substantial enhancement of the $\phi$ meson production in heavy-ion collisions, which was particularly pronounced at sub- and near-threshold energies. 
Interestingly, these findings helped explain the experimentally observed strong enhancement of the $\phi/K^-$ ratio at low energies. 
Unlike alternative approaches, that invoke hypothetical decays of heavy baryonic resonances to the $\phi$ \cite{Steinheimer:2015sha,Steinberg:2018jvv}, our model naturally accounted for this enhancement.

The goal of the present  study is to apply this updated PHSD approach to 12 GeV pA collisions, partially studied at the KEK E325 experiment. This is done with the purpose of gaining a better understanding of the dominant $\phi$ meson production mechanisms, the momentum dependent baryon density profile, that the $\phi$ experiences as it travels through the target nucleus, and  finally the resulting dilepton spectrum generated from its dilepton decay for different in-medium modification scenarios. 
In this paper, we will focus on the theoretical background and provide a detailed description of the studied pA collisions  from the point of view of studying the $\phi$ meson behavior in nuclear matter. 
A reanalysis of the experimental effects, that are specific to the KEK E325 experiment, and subsequent comparisons with the  respective experimental data will be provided in future work.

This paper is organized as follows. In Section \ref{sec:Form}, we give a description of the PHSD 
transport approach used in this work, especially focusing on the treatment of the $\phi$ meson and the parametrization of its possible
in-medium modifications via a density dependent spectral function. 
In Section \ref{sec:results}, we discuss the $\phi$ meson dynamics encountered in the studied pA collisions in detail. 
This includes the kinematical distributions of the $\phi$ meson at the initial stage of the reaction for different production 
channels and the final dilepton spectrum for different in-medium modification scenarios. 
The paper is concluded in Section \ref{sec:Con}.

\section{Formalism \label{sec:Form}}

\subsection{The off-shell PHSD transport approach \label{sec:PHSD}}

Matter produced in relativistic proton-nucleus collisions is strongly interacting and 
hence can be expected to exhibit a behavior distinct from a simple free or weakly interacting gas. 
Therefore a transport approach, which can deal with such a system, is indispensable to study its properties. 
We - for this reason - base the present study of $\phi$ meson production in pA collisions on the Parton-Hadron-String Dynamics (PHSD) 
approach~\cite{Cassing:2008sv,Cassing:2008nn,Cassing:2009vt,Bratkovskaya:2011wp,Linnyk:2015rco,Moreau:2019vhw},
which is a microscopic covariant transport framework used for the dynamical description of strongly interacting hadronic and partonic matter created in heavy-ion collisions. 
It is based on a solution of the Cassing-Juchem generalized off-shell transport equations for test particles \cite{Cassing:1999wx,Cassing:1999mh} derived from the first-order gradient expansion of the Kadanoff-Baym equations \cite{KadanoffBaym} (see Refs.\,\cite{Juchem:2004cs,Cassing:2008nn} for details). 
As it propagates Green's functions (in phase-space representation) that carry information not only on the occupation probabilities (in terms of phase-space distribution functions), but also on the properties of hadronic and partonic degrees of freedom through their spectral functions, we note here that PHSD differs from the semi-classical BUU model. 

The PHSD approach can consistently describe the entire evolution of a relativistic heavy-ion collision, 
encompassing all stages from the initial hard nucleon-nucleon scatterings and string formation, through 
the dynamical deconfinement phase transition to the quark-gluon plasma (QGP) \cite{Berrehrah:2016vzw}, and finally to hadronization and subsequent hadronic interactions in the expanding hadronic phase. 
Notably, PHSD extends the earlier Hadron-String-Dynamics (HSD) transport approach \cite{Cassing:1999es} by incorporating in-medium effects, such as collisional broadening of vector meson spectral functions \cite{Bratkovskaya:2007jk} and modifications of strange degrees-of-freedom based on G-matrix calculations \cite{Cassing:2003vz,Song:2020clw}. 
Furthermore, detailed balance on the level of  $2\leftrightarrow 3$ 
reactions is implemented for the main channels of strangeness production/absorption 
by baryons ($B=N, \Delta, Y$) and pions \cite{Song:2020clw}, and moreover 
for multi-meson fusion reactions $2\leftrightarrow n$ involved in the 
formation of $B +\bar B$  pairs \cite{Cassing:2001ds,Seifert:2017oyb}. 

The multi-particle production in elementary baryon-baryon ($BB$) reactions above the invariant energy $\sqrt{s_{th}^{BB}} =2.65$ GeV, meson-baryon ($mB$) reactions above $\sqrt{s_{th}^{mB}} =2.4$ GeV and meson-meson ($mm$) reactions above $\sqrt{s_{th}^{mm}} =1.3$ GeV  are realized according to the Lund string model~\cite{Nilsson-Almqvist:1986ast} via the event generators FRITIOF 7.02~\cite{Nilsson-Almqvist:1986ast,Andersson:1992iq} and PYTHIA 7.4~\cite{Sjostrand:2006za}. Both generators are ``tuned'' for a better description of the experimental data for elementary $pp$ collisions at intermediate energies~\cite{Kireyeu:2020wou}.

The PHSD model provides a consistent description of the production of hadrons, as well as of electromagnetic probes (dileptons and photons), in p+p, p+A, and A+A collisions from SIS18 to LHC energies (cf. \cite{Bratkovskaya:2013vx,Linnyk:2015rco,Song:2018xca}).

Since in this work we investigate proton-nucleus reactions with relatively small incoming proton  momentum of 12 GeV$/c$, partons and their dynamics are not expected to play any significant role and can be safely neglected. 
We hence use PHSD in its HSD implementation that explicitly treats only hadronic degrees of freedom and their interactions.

\subsection{Modeling of the $\phi$-meson dynamics in PHSD \label{sec:phi_PHSD}}

A detailed discussion of the $\phi$-meson dynamics within PHSD, including off-shell propagation of the in-medium spectral functions, in-medium modified production cross sections, decay to the off-shell strange $K$ and $\bar K$ mesons, is given in Ref.\,\cite{Song:2022jcj}. 
Here, we briefly recall the main aspects of the $\phi$-meson treatment in PHSD, relevant for our present study on $\phi$ dynamics in pA reactions at 12 GeV$/c$.

i) In-medium modifications of $\phi$ mesons \\
In Ref.\,\cite{Song:2022jcj} only the ``collisional broadening scenario" with an increasing $\phi$ meson width, proportional to the surrounding baryon density, was investigated. 
In this work, we explore a wider range of scenarios, including the ``dropping mass scenario", in which the $\phi$-meson pole mass changes at finite baryon density, following the initial Hatsuda/Lee \cite{Hatsuda:1991ez} prediction (see also Refs.\,\cite{Gubler:2014pta,Kim:2019ybi} for later updates), which has attracted new attention due to the ALICE measurements \cite{ALICE:2021cpv} and lattice QCD calculations \cite{Lyu:2022imf} of the $\phi N$ scattering length. 

It should be noted here that the modification of the $\phi$ meson mass in a dropping mass and/or collisional broadening scenario leads to the lowering of the $\phi$ meson production threshold in all hadronic reactions in the nuclear medium. 
This then enhances the $\phi$ meson production at subthreshold energies and hence modifies both elastic and inelastic hadronic scattering cross sections. This effect is significant particularly at subthreshold energies realized in low energy heavy-ion collisions as studied in Ref.\,\cite{Song:2022jcj}.

ii) Novel  $mB \to \phi$ production channels.\\
In this study we make use of the framework discussed in Ref.~\cite{Song:2022jcj}, in which novel meson-baryon and meson-hyperon production channels for the $\phi$ meson, obtained from a T-matrix coupled channel approach based on an extended SU(6) chiral effective model, have been implemented for the first time within a transport approach. 
In that work, the SU(6) chiral effective T-matrix approach was specifically used to calculate the different meson-baryon $\phi$ meson production cross sections. 

Including the five sets of total quantum numbers $(I,S,J)=(1/2,0,1/2)$, $(3/2,0,1/2)$, $(1/2,0,3/2)$, $(3/2,0,2/2)$, $(3/2,0,5/2)$, 10 scattering channels to produce $\phi+N$ and 7 channels to produce $\phi+\Delta$ are taken into account, extending the usually considered $m+B$ channels involving pions, nucleons and $\Delta$'s by accounting for s-wave scattering $m+B \to \phi + B$ reactions with $m=\eta, K, \rho, K^*$ mesons and $B=N, \Delta, \Lambda, \Sigma, \Sigma^*$ baryons for all possible $I=1/2$ and 3/2 channels. As was shown in Ref.~\cite{Song:2022jcj}, the $\eta N \to \phi N$ channel turned out to be the most important, even though the corresponding cross section is not the largest one compared to other channels. This can be understood from the $\eta$ abundance, which is larger than that for all other mesons (besides the pion), making the above mentioned channel important at SIS energies. 
We emphasize here again that most of these $mB$ channels so far have been neglected in transport studies.

iii)  The in-medium modifications of $K$ and $\bar K$ mesons. \\
Since the $\phi$ meson strongly couples to the open strange mesons $K,\bar{K}$, it becomes critical to take their in-medium modifications properly into account. Following the previous PHSD study of Ref.~\cite{Song:2020clw}, in this work we include the in-medium modifications of strange mesons. 
Specifically, we assume a repulsive in-medium potential for kaons which increases linearly with baryon density, leading to an increase of the kaon mass with increasing density while its width remains negligibly small. For anti-kaons, which have a more complicated in-medium behavior because of the attactive $\bar{K}N$ interaction and the ensuing generation of the $\Lambda(1405)$ resonance, we employ the density and temperature dependent complex self-energy obtained from a G-matrix approach. This leads to a substantial broadening of the corresponding spectral function and to a decrease of the pole mass of $\bar K$ in dense matter. 

\subsection{Modeling of the $\phi$ meson spectral function \label{sec:phi_spec}}

The properties of the $\phi$ meson in the PHSD framework are expressed in terms of a Breit-Wigner spectral function 
\begin{align}
A_{\phi}(M, \rho) =  C \frac{2}{\pi} 
\frac{M^2 \Gamma^{\ast}_{\phi}(M, \rho)}{[M^2 - M^{\ast 2}(\rho)]^2 + M^2 \Gamma^{\ast 2}_{\phi}(M, \rho)}, 
\label{eq:spec_func}
\end{align}
where $C$ is a renormalization constant determined from the condition 
\begin{align}
\int_{M_{\mathrm{min}}}^{M_{\mathrm{max}}} A_{\phi}(M, \rho) dM = 1,  
\label{eq:normalization}
\end{align}
in which $M_{\mathrm{min}} = 3 m_{\pi}$ for the vacuum case ($\rho = 0$) 
and $M_{\mathrm{min}} = 2 m_{e}$ for 
in-medium scenarios at finite density; 
$M_{\mathrm{max}}$ is set to 2 GeV for all considered cases. 

\subsubsection{Dropping pole mass scenario}

The density dependent mass $M^{\ast}(\rho)$ is parametrized as 
\begin{align}
M^{\ast}(\rho) = \frac{M_0}{1 + \alpha \rho/\rho_0}, 
\label{eq:mass_density}
\end{align}
where $M_0$ is the $\phi$ meson vacuum mass, $\rho_0$ stands for the normal nuclear matter density 
and the parameter $\alpha$ determines the magnitude of the mass shift. 
Note that for the relatively small mass shifts considered for the $\phi$ meson at densities up to and 
around $\rho_0$ in this work, Eq.\,(\ref{eq:mass_density}) is for all practical purposes equivalent to the 
more conventional linear parametrization of $M^{\ast}(\rho) = M_0 (1 - \alpha \rho/\rho_0)$ 
in line with the Hatsuda/Lee \cite{Hatsuda:1991ez} or Brown/Rho scaling \cite{Brown:1991kk}.

\subsubsection{Collisional broadening scenario}

The density dependence of the width $\Gamma^{\ast}_{\phi}(M, \rho)$ is given as 
\begin{align}
\Gamma^{\ast}_{\phi}(M, \rho) &= \Gamma_{\phi}(M) + \alpha_{\mathrm{coll}} \frac{\rho}{\rho_0},
\label{eq:width_density}
\end{align}
where $\Gamma_{\phi}(M)$ is the $\phi$ vacuum decay width  
including $\phi\to\rho\pi(3\pi)$, calculated using an effective chiral Lagrangian approach with parity anomalous terms~\cite{Kaymakcalan:1983qq} - cf. Section II in Ref. \cite{Song:2022jcj}, while the second term represents the collisional width broadening due to finite density effects. 

We note, that the linear density terms of both mass and width in Eqs.\,(\ref{eq:mass_density}) and (\ref{eq:width_density}) can in principle depend on the particle momentum (see Refs.\,\cite{Lee:1997zta,Kim:2019ybi} and 
Refs.\,\cite{Cabrera:2003wb,Vujanovic:2009wr,Cabrera:2017agk} 
for calculations of such effects using QCD sum rules and hadronic effective theories, respectively). We ignore this momentum dependence in this work, which considerably simplifies the particle equations of motions \cite{Bratkovskaya:2007jk}. 
It may, however, become important to take such effects into account when analyzing the data of the 
future J-PARC E16 experiment, which are expected to be precise enough to study the 
detailed momentum dependence of the dilepton spectrum from pA reactions \cite{Ashikaga:2019jpc}. 

\subsubsection{Compilation of all considered scenarios}

\begin{figure}[!h]
\includegraphics[width=8.0cm]{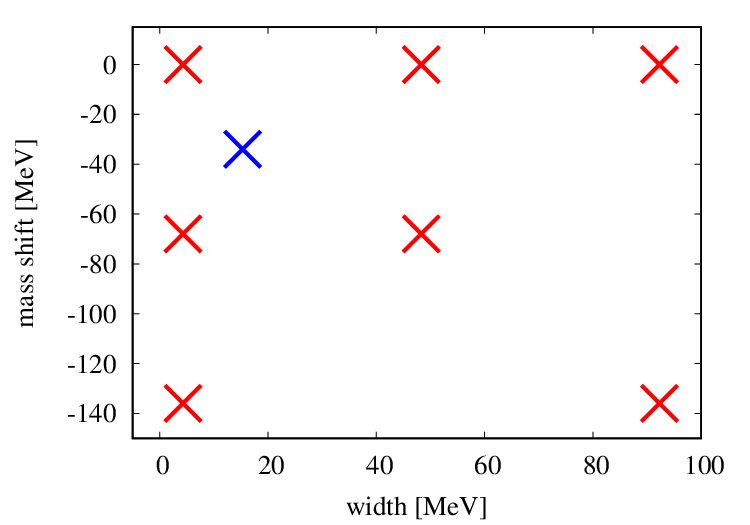}
\caption{Modification scenarios for the $\phi$-meson mass shift and width broadening at normal nuclear matter density, considered in this work (red crosses). 
The blue cross indicates the location corresponding to the value reported by the KEK E325 collaboration \cite{KEK-PS-E325:2005wbm}.}
\label{fig:scen}
\end{figure}

For the parameters $\alpha$ and $\alpha_{\mathrm{coll}}$ governing the density dependence of the mass 
and width in Eqs.\,(\ref{eq:mass_density}) and (\ref{eq:width_density}), we employ multiple combinations of 
values, which correspond to scenarios with different degrees of mass shifts and broadenings. 
These are shown in Fig.\,\ref{fig:scen} in terms of mass shift and width values at $\rho_0$. 

\begin{figure}
\includegraphics[width=8.0cm]{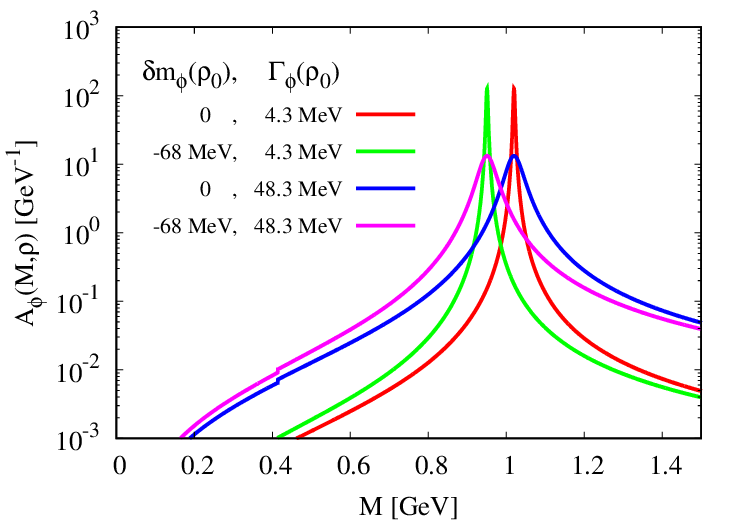} 
\caption{The $\phi$ meson spectral function $A_{\phi}(M, \rho)$ for a density independent scenario (red line), and representative dropping pole mass (green line), collisional broadening (blue line) and mixed (purple line) scenarios at normal nuclear matter density $\rho= \rho_0$.}
\label{fig:SF}
\end{figure}

The corresponding spectral functions are depicted in Fig.\,\ref{fig:SF}, where both the vacuum ($\rho= 0$)  
and nuclear matter ($\rho = \rho_0$) cases are shown. 

\section{Results \label{sec:results}}

\subsection{$\phi$ meson production on C, Cu and Pb targets \label{sec:phi_production}}

\begin{figure*}
\includegraphics[width=5.8cm]{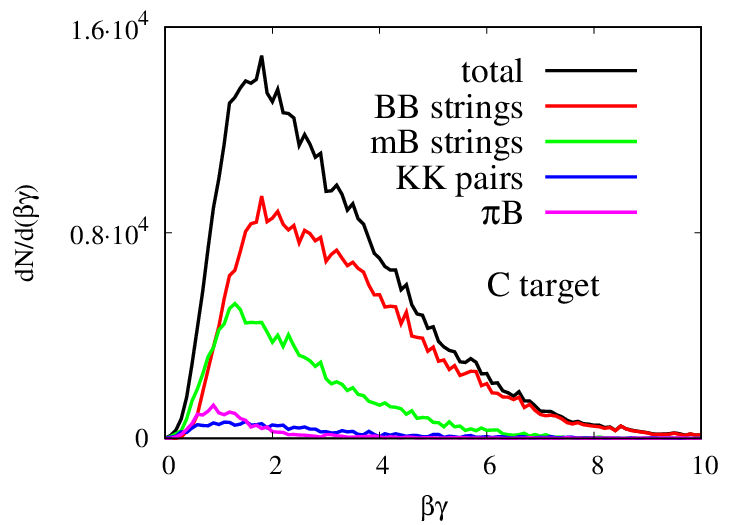}
\includegraphics[width=5.8cm]{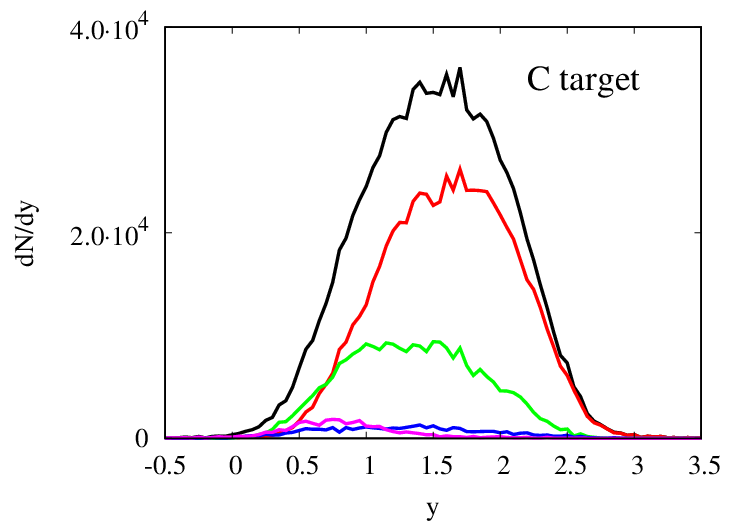}
\includegraphics[width=5.8cm]{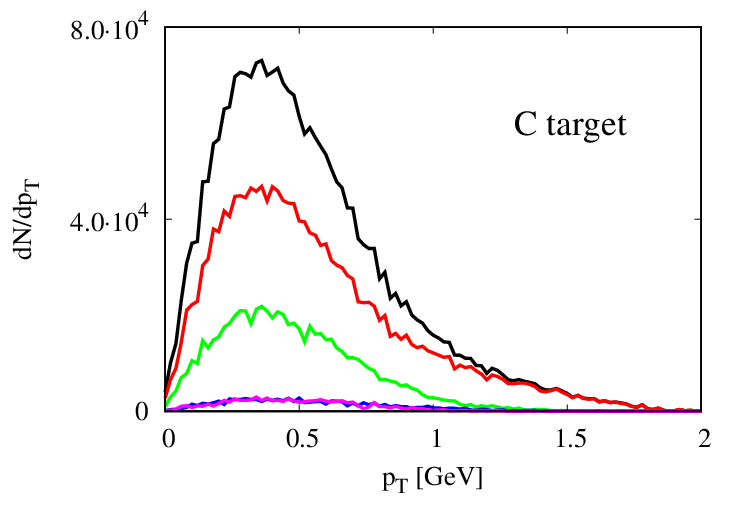} \\
\includegraphics[width=5.8cm]{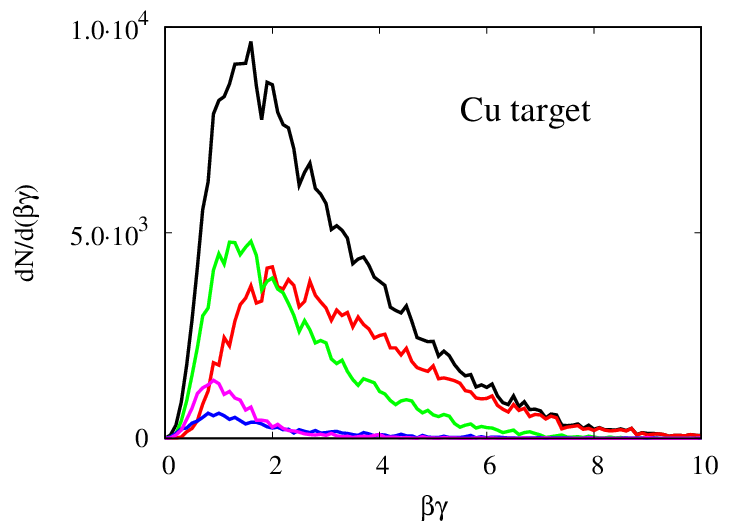}
\includegraphics[width=5.8cm]{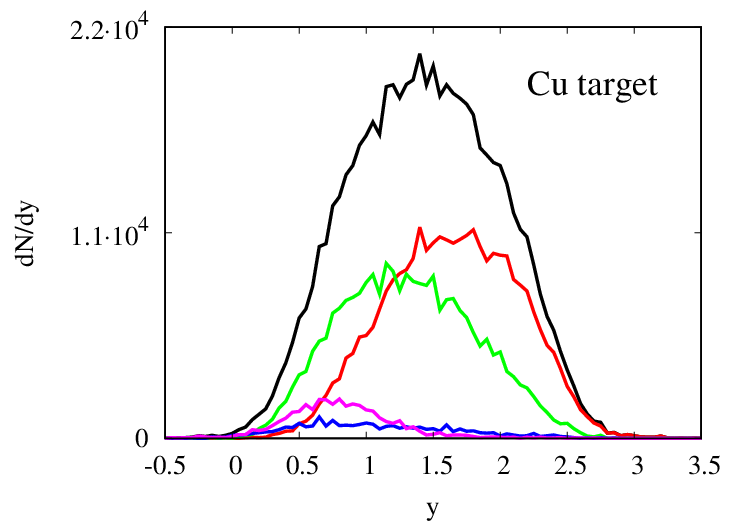}
\includegraphics[width=5.8cm]{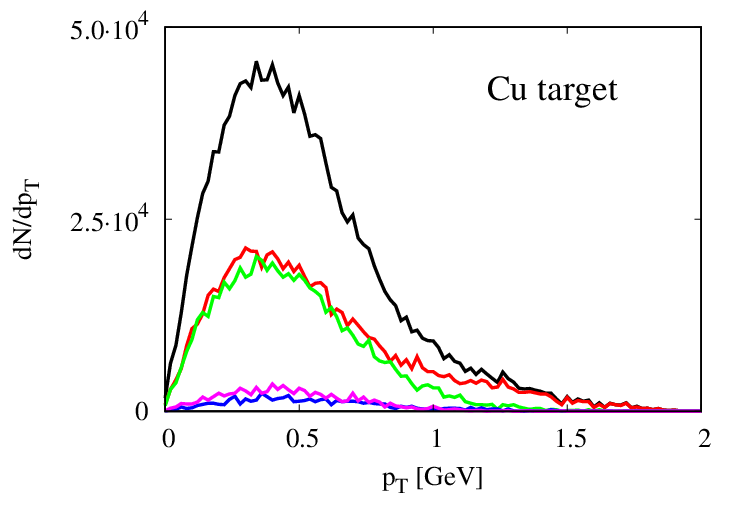} \\
\includegraphics[width=5.8cm]{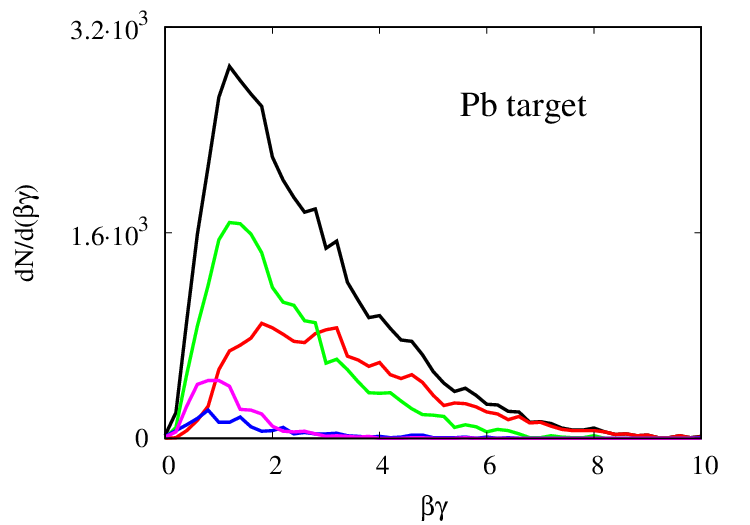}
\includegraphics[width=5.8cm]{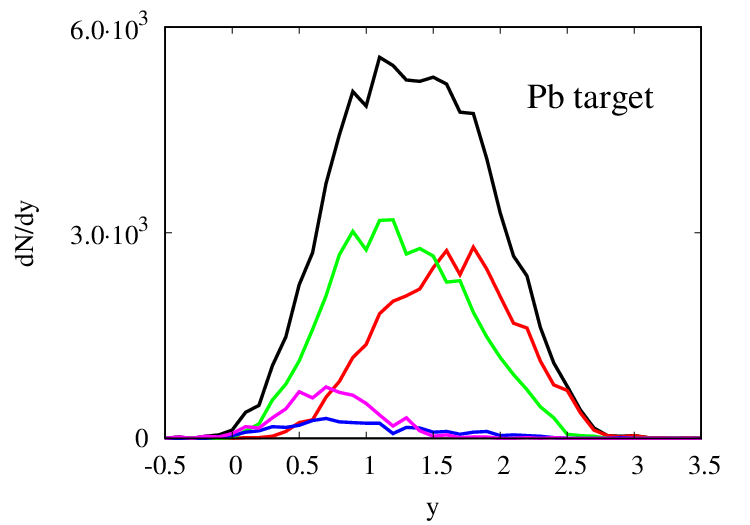}
\includegraphics[width=5.8cm]{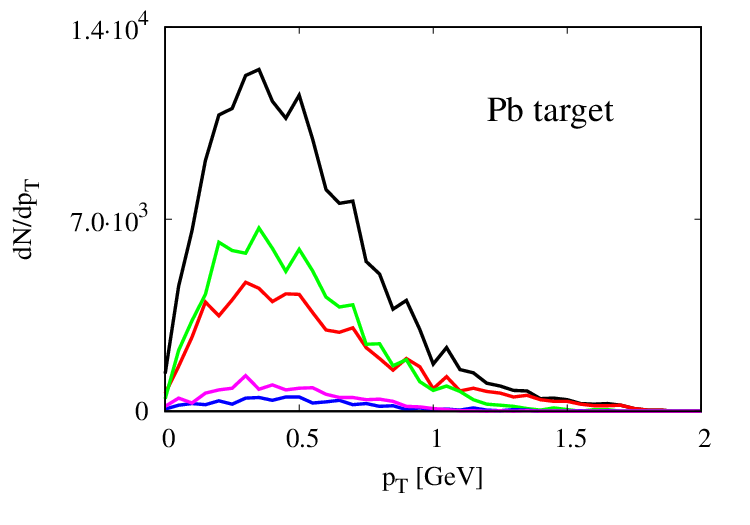}
\vspace{-0.2cm}
\caption{$\beta \gamma$ (left), rapidity (middle) and $p_{\mathrm{T}}$ (right) distributions of the generated $\phi$ mesons at their 
production time for carbon (top), copper (middle) and lead (bottom) targets. 
The coloured lines show the different production channels: ``BB (mB) strings" stand for the $\phi$ production by strings generated in 
high-energy baryon-baryon (meson-baryon) collisions [red (green) lines], 
``$KK$ pairs" indicate $\phi$ production by $K\bar K$ 
reactions (blue lines) and 
``$\pi B$" show $\phi$s produced by low energy $\pi B$ collisions (purple lines), both below the string threshold, while ``total" stands for the total number of $\phi$s from all possible channels (black lines).
}
\label{fig:prod_A}
\end{figure*}

We first examine the kinematical distributions of the $\phi$ meson at the time of its formation in some 
detail. The generated $\beta \gamma$, rapidity and $p_{\mathrm{T}}$ distributions 
of the $\phi$ meson on C, Cu and Pb targets at the respective production time are shown in Fig.~\ref{fig:prod_A}. 
In the same figures, the contributions of the various production mechanisms to the 
different distributions are also illustrated. 

As seen in Fig.\,\ref{fig:prod_A}, 
the $\phi$ mesons with the highest momenta are predominantly produced
in the initial high-energy baryon-baryon collisions by string formation. On the other hand, $\phi$ mesons that are generated from strings produced in meson-baryon collisions, that occur at a later stage of the simulated reaction, tend to be distributed at smaller momenta. Hadronic generation mechanisms, such as $K\bar{K}$ pair annihilation or inelastic low energy $\pi$-baryon scattering, produce $\phi$s of significant numbers only at relatively small momenta. 
The above description roughly holds for all targets, the most significant difference between them 
being the relative numbers of $\phi$ mesons produced in baryon-baryon (BB) versus meson-baryon (mB) strings. While for the lightest C target, BB dominates over mB, the situation is reversed when going to the heavier targets, especially for the Pb case. 
This behavior can be understood from the secondary nature of the mB strings. Produced energetic mesons, that will further collide with baryons to form mB strings if the medium is large enough, are more likely to leave the dense region in case of the light C target.   

\begin{figure*}
\includegraphics[width=11.0cm]{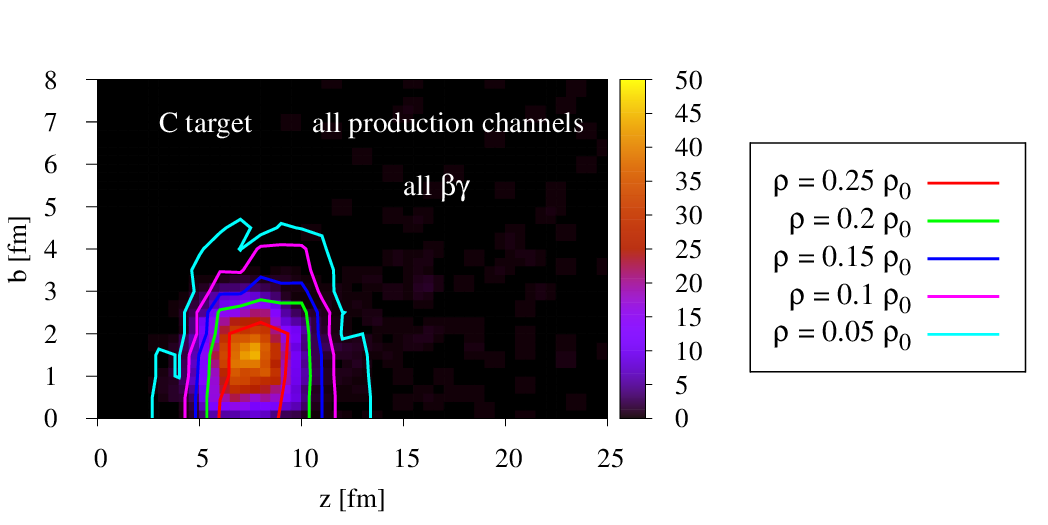} \\
\vspace*{-0.3cm}
\includegraphics[width=7.5cm]{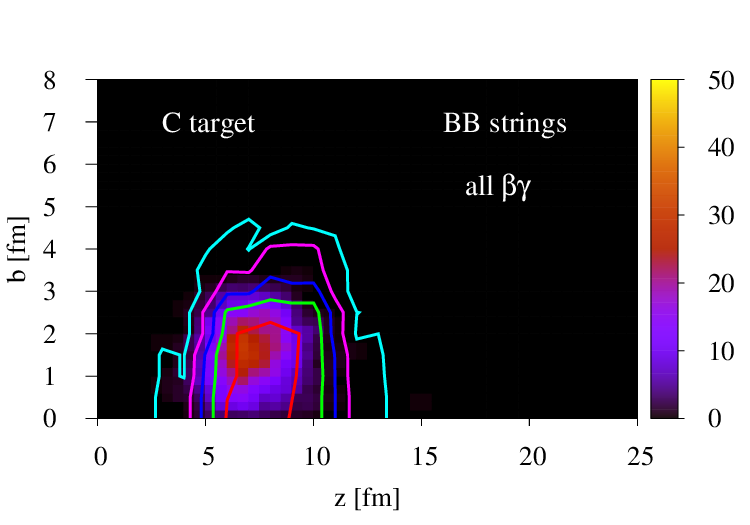}
\includegraphics[width=7.5cm]{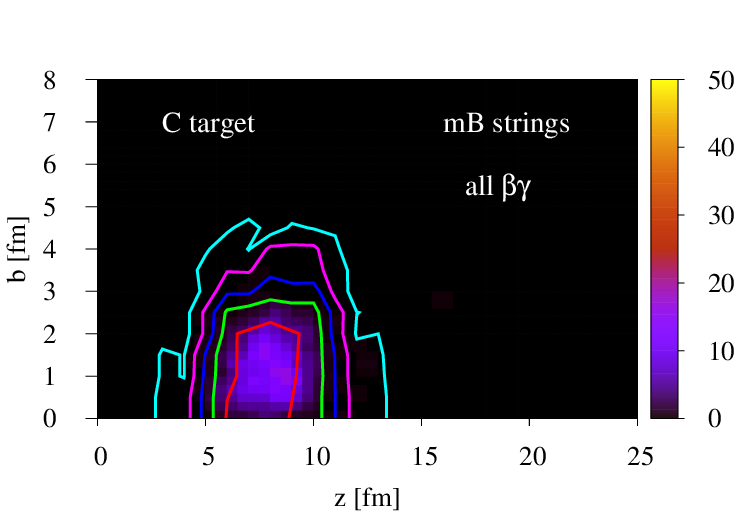} \\
\includegraphics[width=7.5cm]{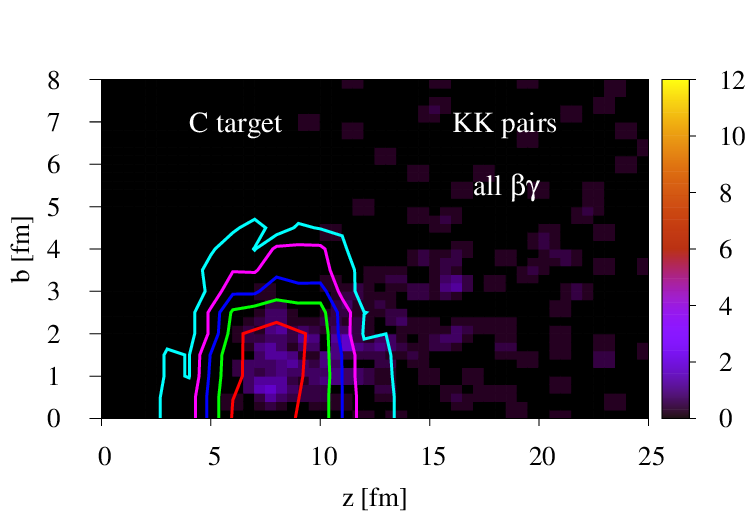}
\includegraphics[width=7.5cm]{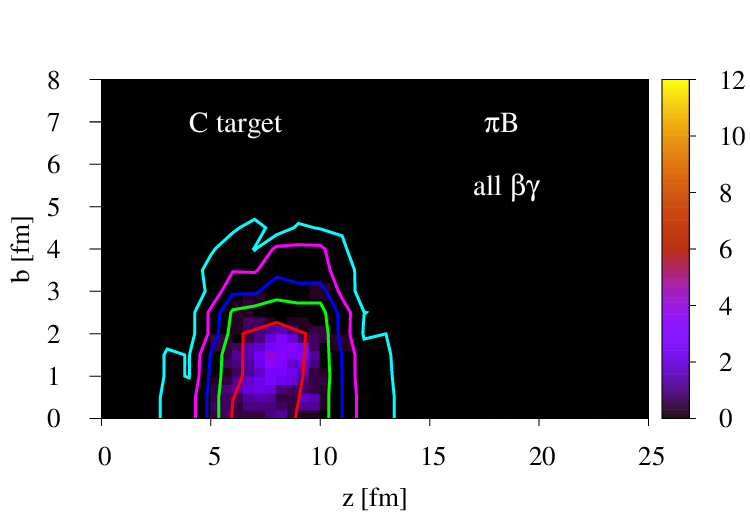} 
\caption{Spatial distributions of the total $\phi$ meson production (top plot), and the respective distributions subdivided into 
the different production mechanisms. The colored lines stand for the averaged densities of the target carbon nucleus in the lab frame. 
Note that, because the hadronic production mechanisms, shown in the two bottom plots, are much less frequent, we have changed 
the color scale for these two plots to improve their visibility.}
\label{fig:prod_loc_C}
\end{figure*}

\begin{figure*}
\includegraphics[width=11.0cm]{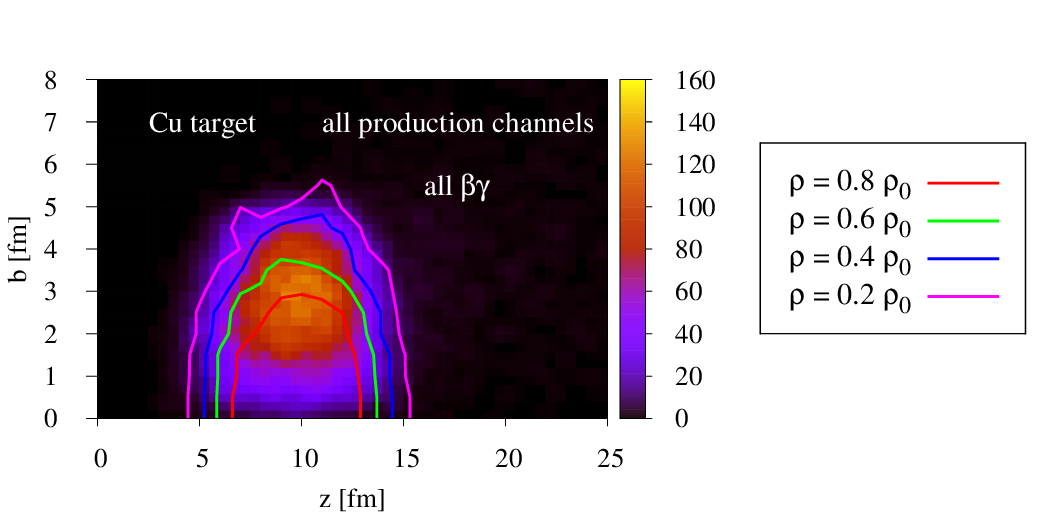} \\
\vspace*{-0.3cm}
\includegraphics[width=7.5cm]{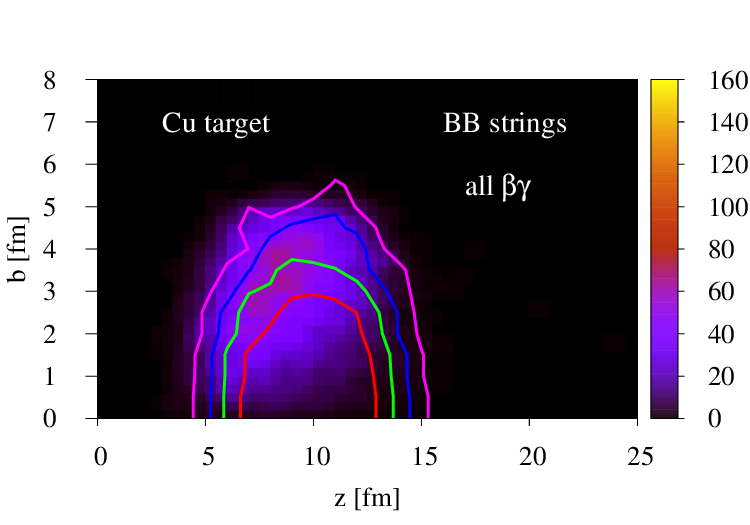}
\includegraphics[width=7.5cm]{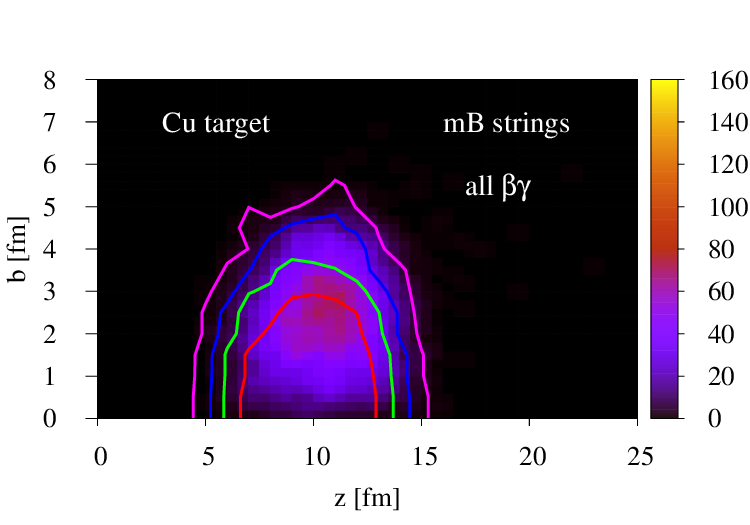} \\
\includegraphics[width=7.5cm]{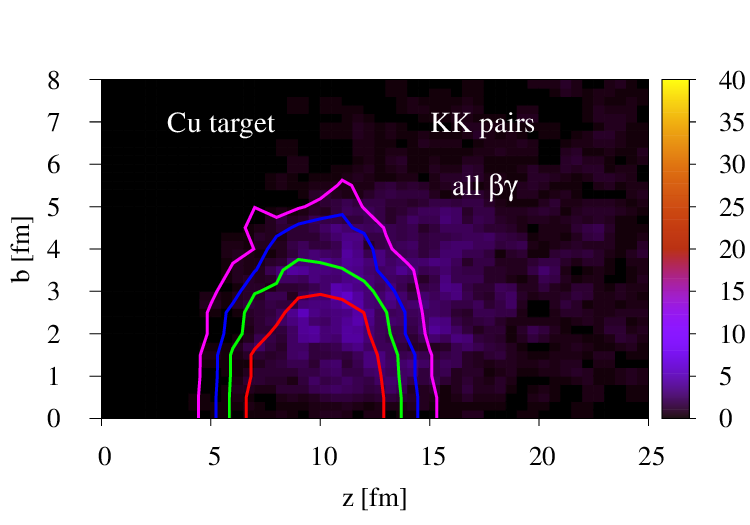}
\includegraphics[width=7.5cm]{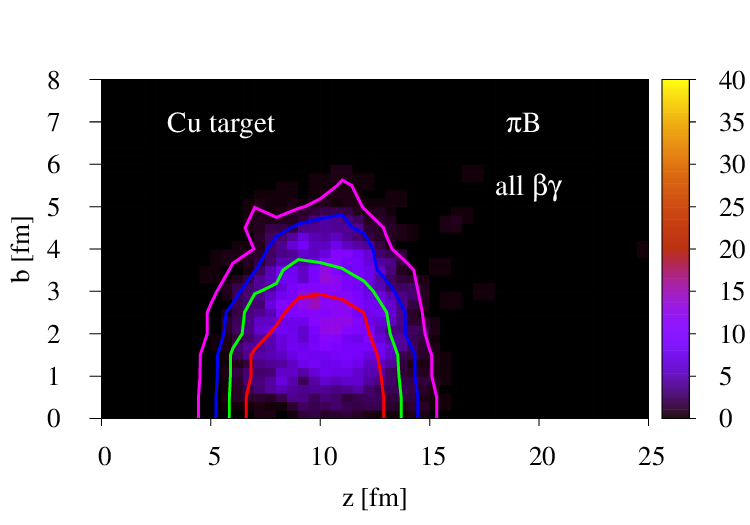} 
\caption{Same as Fig.\,\ref{fig:prod_loc_C}, but for the copper target.}
\label{fig:prod_loc_Cu}
\end{figure*}

\begin{figure*}
\includegraphics[width=11.0cm]{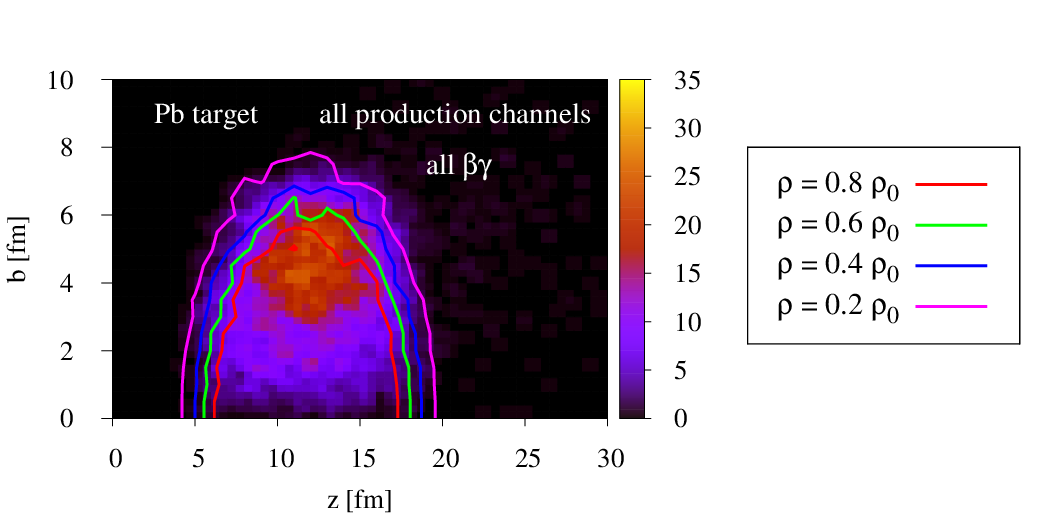} \\
\vspace*{-0.3cm}
\includegraphics[width=7.5cm]{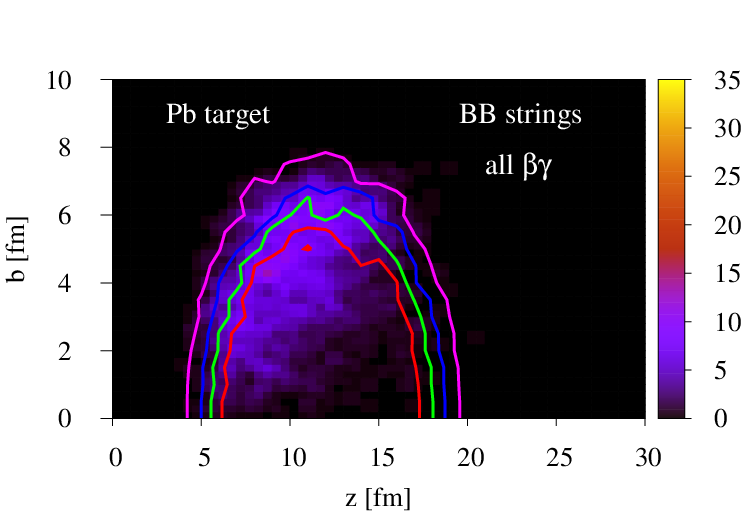}
\includegraphics[width=7.5cm]{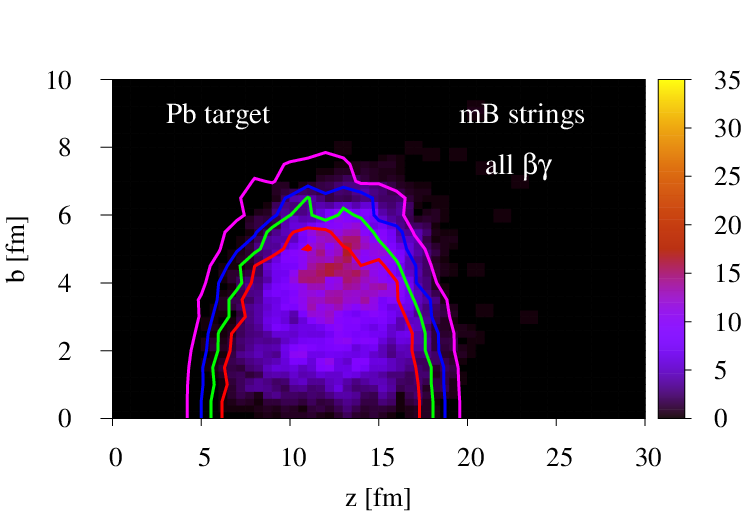} \\
\includegraphics[width=7.5cm]{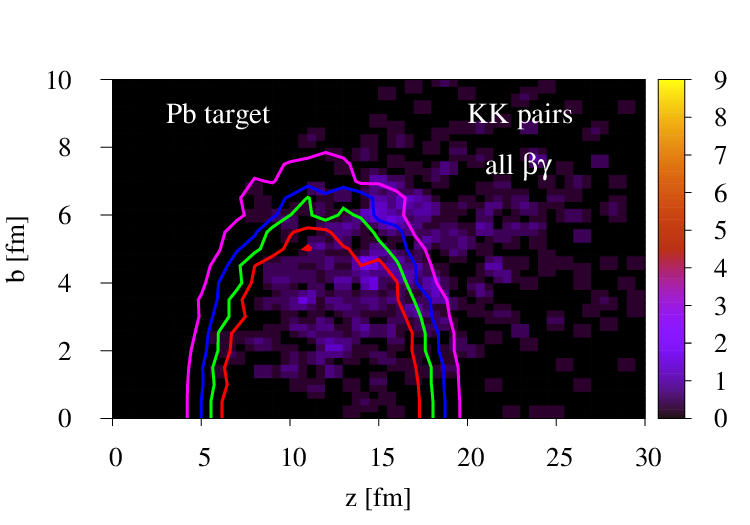}
\includegraphics[width=7.5cm]{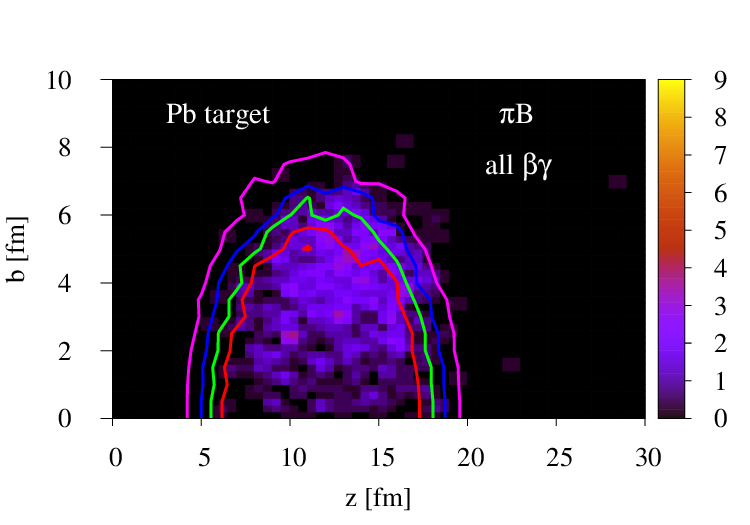} 
\caption{Same as Fig.\,\ref{fig:prod_loc_C}, but for the lead target.}
\label{fig:prod_loc_Pb}
\end{figure*}

Next, we show in Figs.\,\ref{fig:prod_loc_C}, \ref{fig:prod_loc_Cu} and \ref{fig:prod_loc_Pb} where and how the $\phi$ mesons are produced in the different target nuclei. For this, we follow 
Ref.\,\cite{Bratkovskaya:2001ce} and define the radius $b = \sqrt{x^2 + y^2}$ of a circle perpendicular to the beam, which is aligned with the z-axis, the incoming proton approaching the target from the left. 
In Figs.\,\ref{fig:prod_loc_C}, \ref{fig:prod_loc_Cu} and \ref{fig:prod_loc_Pb}, we display the relative numbers produced in each cell, $\frac{dN}{dbdz}$, 
The top plots in each figure show the total number of produced $\phi$ mesons, while the 
four plots below give the contributing numbers of the different production mechanisms. Note 
that the meaning of the gray scale measure is different for each plot. 
The solid lines correspond to the averaged baryon densities of the target nucleus in the lab frame. 

For the carbon target of Fig.\,\ref{fig:prod_loc_C}, it is clear that due to its relatively small size 
the density of the ``medium" reaches at most 0.25 $\rho_0$, meaning that the dileptons produced 
from such a p-C reaction will not be very sensitive to any density-effects. 
Nevertheless, a tendency of the $\phi$ mesons originating from BB-strings to be produced on 
the left side of the target is already visible. Furthermore, in contrast to all other production mechanisms, 
a large fraction of $\phi$'s produced via kaon pair annihilation, are produced outside of the target. 
For the copper and lead targets (Figs.\,\ref{fig:prod_loc_Cu} and \ref{fig:prod_loc_Pb}), it is interesting 
to observe that, while the total spatial production distribution is roughly symmetric with respect to the target center, it can be decomposed into different production mechanisms which characteristically distinct features. 
The production via strings generated by baryon-baryon collisions occurs mainly on the left side of the target, where it is hit first by the incoming proton. Thereafter, the production by secondary meson-baryon strings happens closer to the center of the target. 
The hadronic production channels, that become dominant at the latest stage of the collision, also exhibit a different behavior. Kaon pairs produce $\phi$ mesons both inside and outside of the target nucleus, while the production via inelastic $\pi$-baryon scattering naturally occurs mostly inside of the dense target region, where many nucleons are available for scattering. Note that since 
head-on collisions rarely happen (because of the geometrical factor $2 \pi b$), the distribution is suppressed for small $b$ values.

\subsection{The dilepton spectrum in the $\phi$ meson mass region \label{sec:dil_phi_mass}}

\begin{figure*}
\includegraphics[width=5.8cm]{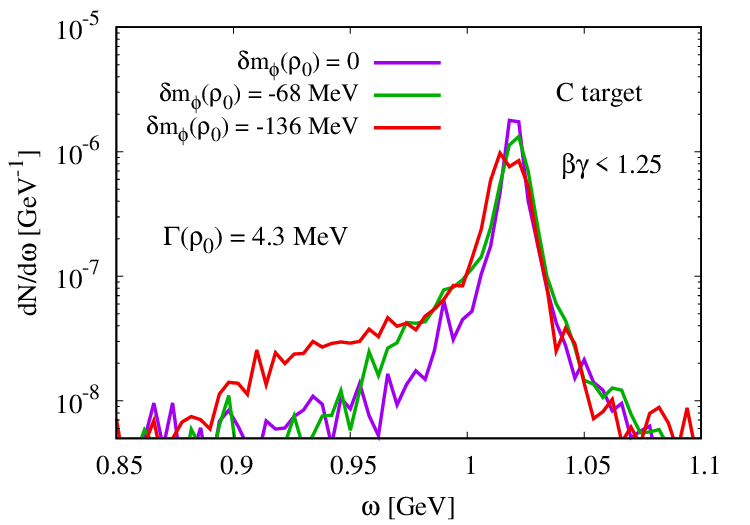}
\includegraphics[width=5.8cm]{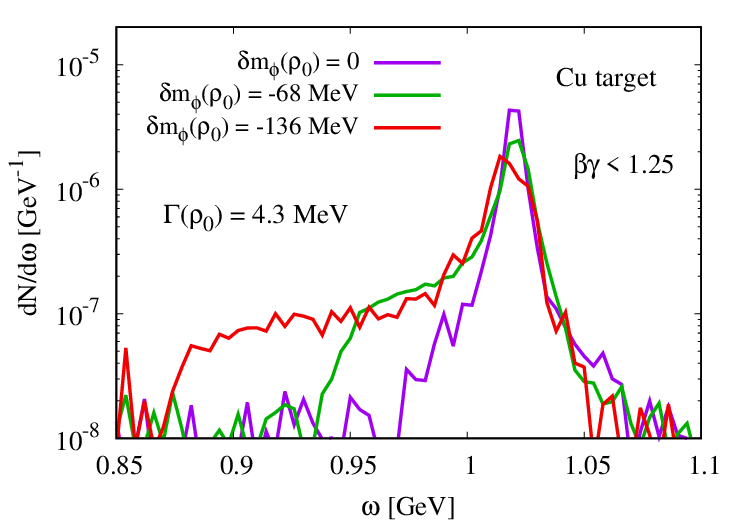}
\includegraphics[width=5.8cm]{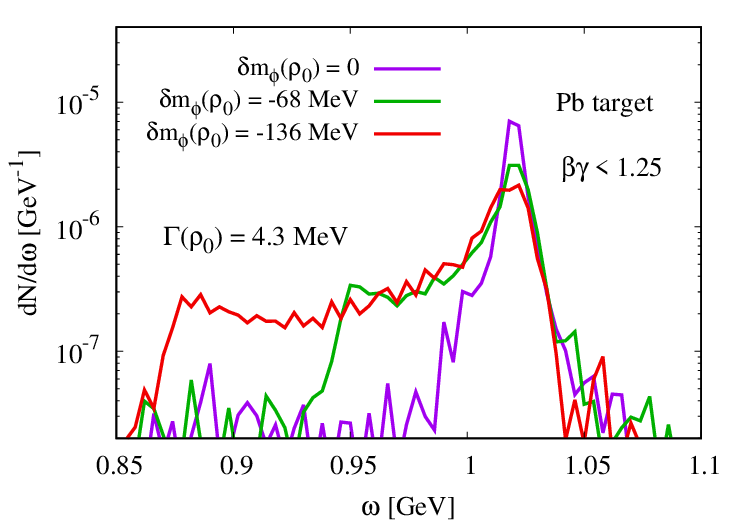}
\includegraphics[width=5.8cm]{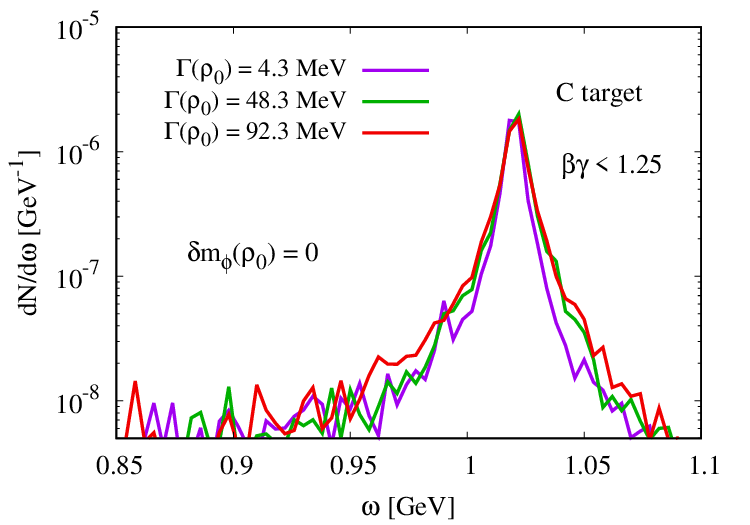}
\includegraphics[width=5.8cm]{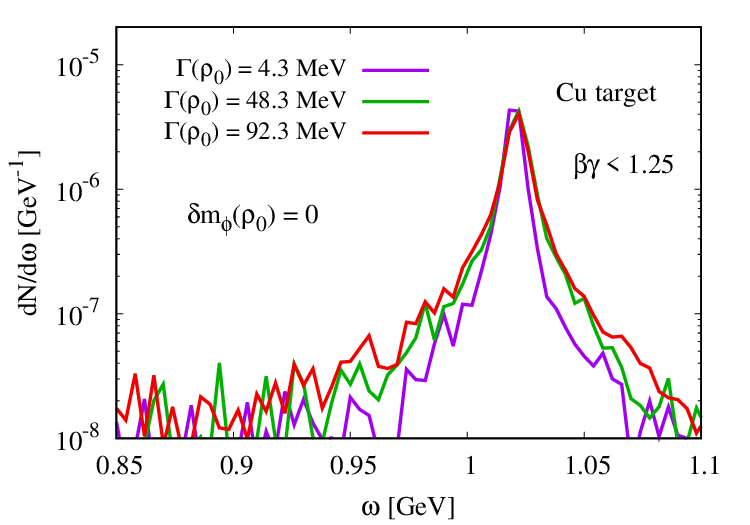}
\includegraphics[width=5.8cm]{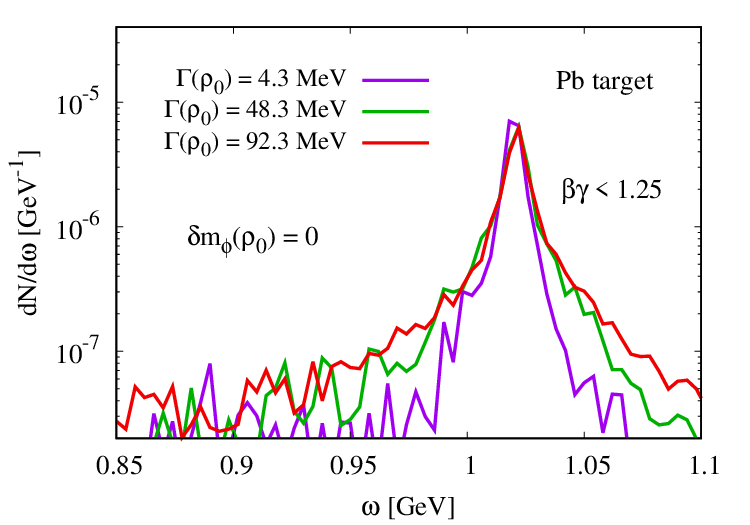}
\includegraphics[width=5.8cm]{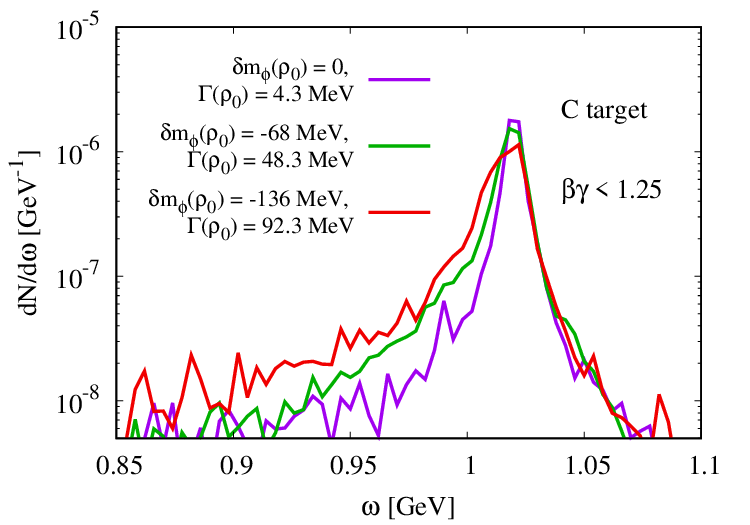}
\includegraphics[width=5.8cm]{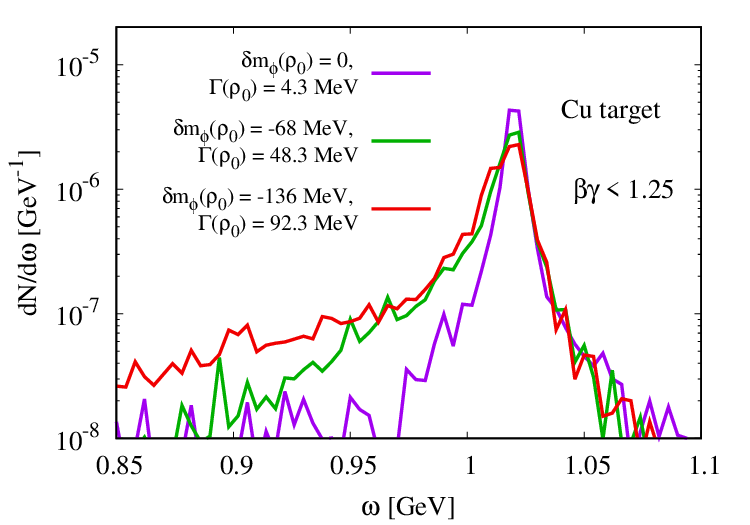}
\includegraphics[width=5.8cm]{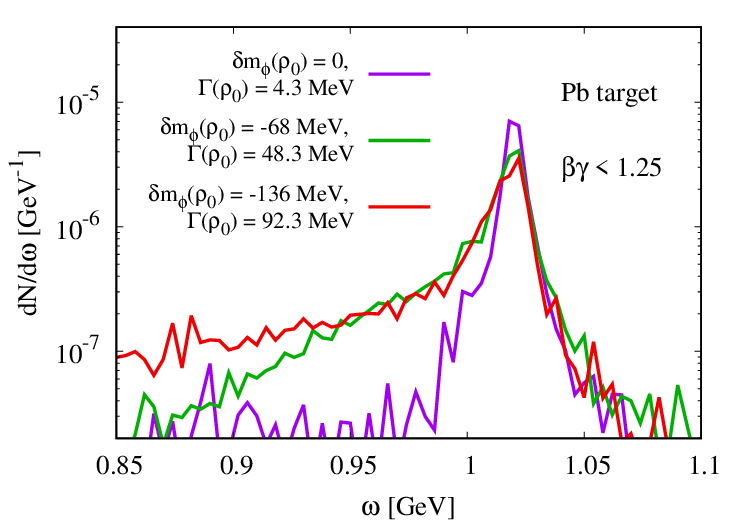}
\caption{The dilepton spectrum in the $\phi$ meson mass region obtained in our transport simulation of 
pA reactions with a C target (left plots) and Cu target (right plots), for 
various finite density modification scenarios of the $\phi$ meson spectral function. 
Pure mass shift, broadening and mixed scenarios are respectively shown in the top, middle and bottom plots.}
\label{fig:dilepton_phi_mass}
\end{figure*}

Let us next compare the dilepton spectra obtained in the $\phi$ meson mass region for a few representative examples of the in-medium modification scenarios shown in Fig.\,\ref{fig:scen}. In Fig.\,\ref{fig:dilepton_phi_mass},  we display the spectra  for a pure mass shift (top plots), broadening (middle plots) and a combination of the two (bottom plots), for  C (left plots), Cu (middle plots) and Pb targets (right plots). 

We have here limited the lab-frame momenta of the dileptons to $\beta \gamma < 1.25$, 
which represents a momentum range that is both experimentally 
accessible (it corresponds to the lowest $\beta\gamma$ bin of the KEK E325 experiment \cite{Muto:2007iks}), 
and where one expects to find the largest density effects, because for slow momenta the $\phi$ mesons 
have more time to interact with the nuclear medium before they leave the target. 

The figures show that, for the scenarios considered here, mass shifts have a more significant effect 
and generate a shoulder on the left side of the peak which is more or less pronounced 
depending on the magnitude of the mass shift. The main peak at the original location of the 
$\phi$ meson vacuum mass, which is generated by the particles decaying outside of the dense medium, 
is present for all scenarios. A double-peak structure \cite{Asakawa:1994nn,Asakawa:1994ez} can only be generated for the 
most extreme cases, with a negative mass shift, no broadening effects and the largest Pb target.   

\section{Conclusions and Outlook \label{sec:Con}}

In this work we have studied 12 GeV pA reactions with C, Cu and Pb targets within the PHSD transport approach, with the goal of determining the effects of the dense nuclear environment on the $\phi$ mesons, which are generated dominantly during the initial stages of the reaction, on their subsequent dilepton decays and the corresponding dilepton spectrum. 
PHSD allows to take into account the modifications of the in-medium vector meson spectral functions 
as the particles move through the nuclear target volume, including off-shell effects, and the 
regeneration of their vacuum form once they leave the dense environment. 
This approach thus makes it possible to study the sensitivity of the (experimentally measurable) 
dilepton spectrum on the different mass shift and broadening modification scenarios of the $\phi$ meson 
in nuclear matter, that have been proposed in the literature. 

We first examined and compared the different $\phi$ meson production mechanisms as a function 
of both kinematics and location in the target. We see that at 12 GeV collision energy studied here, 
the high energy string production generally dominates, with initial baryon-baryon strings dominating 
for the lighter C target, while the secondary meson-baryon strings become the most important production mechanism for the Pb target (see Fig.\,\ref{fig:prod_A}). 
This result can be better understood by looking at the spatial distributions of the different 
mechanisms in and around the target region, that are shown in Figs.\,\ref{fig:prod_loc_C}-\ref{fig:prod_loc_Pb}. 
We see there that baryon-baryon string production happens dominantly on the target surface, where 
it is hit by the incoming projectile. The meson-baryon string production occurs in secondary collisions 
inside or on the rear-side of the target. 
Hadronic production, through $K\bar{K}$ annihilation and low-energy $\pi$-baryon collisions, 
is not completely negligible, but takes place at a rate one order of magnitude smaller compared 
to strings. 

Next, turning to the dilepton spectrum, we find that, even though nuclear absorption reactions cause the $\phi$ meson lifetime to be significantly reduced in nuclear matter, a majority of the $\phi$'s still decay outside of the nuclear target, which limits the magnitude of nuclear modification effects. This is especially true for broadening scenarios of the spectral function, which alter the dilepton spectrum only marginally. On the other hand, mass shift scenarios do leave a more prominent mark on the outgoing dileptons, especially for the heavier  Cu and Pb targets. For the Pb case, if large mass mass shifts in nuclear matter of -68 MeV or -136 MeV and no broadening is assumed, we find that a small second peak at the corresponding reduced mass value can be generated (see top right plot in Fig.\,\ref{fig:dilepton_phi_mass}), which is however washed out once 
broadening effects are taken into account (bottom right plot in Fig.\,\ref{fig:dilepton_phi_mass}). 

It will be interesting to see how our results compare with available experimental data of the 
KEK E325 experiment \cite{KEK-PS-E325:2005wbm}. To properly carry out such a comparison, detailed information 
about the experimental acceptance and resolution of this experiment will however be needed. Furthermore, 
QED effects on the outgoing dileptons, especially through radiative corrections are known to 
enhance the spectrum on the negative side of vector meson peaks \cite{Spiridonov:2004mp,Muto:2007iks}, 
similar to what we have found in this work for mass shift scenarios of the $\phi$ meson. 
An analysis in collaboration with researchers of the KEK E325 experiment, which will take all the above effects 
into account, is currently ongoing, the results of which will be reported in a future publication.

\section*{Acknowledgements}

We are grateful to W. Cassing, V. Metag, L. Tolos, M. Naruki and S. Yokkaichi for inspiring discussions. 
P.G. is supported by the Grant-in-Aid for Scientific Research (A)  (JSPS KAKENHI Grant Number JP22H00122). 
M.I. is supported by the Grant-in-Aid for Scientific Research (S)  (JSPS KAKENHI Grant Numbers JP20H05647 and JP23H05440). 
Furthermore, we acknowledge support by the European Union’s Horizon 2020 research and innovation program under grant agreement No 824093 (STRONG-2020). 
The computational resources have been partially provided by the LOEWE-Center for Scientific Computing and the ``Green Cube" at GSI, Darmstadt.

\bibliography{references_EB} 


%
%
\end{document}